\documentclass[letterpaper,12pt, draft, leqno]{article} % Comments after % are ignored
\usepackage{amsmath,amssymb,amsfonts, amsthm} % Typical maths resource packages
\usepackage[mathscr]{eucal}
\usepackage{color, xcolor} % For creating coloured text and background

\newtheorem{theorem}{Theorem}[section]

\newtheorem{lemma}{Lemma}[section]
\newtheorem{definition}{Definition}[section]
\newtheorem{assumption}{Assumption}[section]
\theoremstyle{definition}
\newtheorem{remark}{Remark}[section] 
\newtheorem{example}{Example}[section]
\numberwithin{equation}{section}

\newcommand{\envspace}{\vspace{2mm}}

\newcommand{\mm}{market maker} 
\newcommand{\Real}{\mathcal{R}}

\newcommand{\PP}{\mathbb{P}}

\newcommand{\FF}{\mathcal{F}}
\newcommand{\HH}{\mathcal{H}}
\newcommand{\EE}{\mathbb{E}}

\newcommand{\half}{\frac{1}{2}} 
\newcommand{\ind}{{\mathbf 1}}
 
\newcommand{\Ec}[2]{\EE\left[\left. #1 \right|  \FF_{#2}\right]} 
\newcommand{\evalepszero}[1]{\left. #1 \right|_{\epsilon=0}} 
\newcommand{\SeH}{S^{\epsilon H}}
\newcommand{\partialepsilon}{\frac{\partial}{\partial\epsilon}}
\newcommand{\variance}[1]{\langle #1 \rangle} 
\newcommand{\brakets}[1]{\left(  #1 \right)}

 \def\keywordname{{\bf Key words:}} 
 \newcommand{\keywords}[1]{\par\addvspace\baselineskip\noindent\keywordname\enspace\ignorespaces #1}

\title{Pricing in an equilibrium based model for a large investor. \thanks{The author is grateful to Dmitry Kramkov for helpful discussions and insightful comments.}}
 
\author{David German \\ \vspace{-2mm} {\small Claremont McKenna College} \\ \vspace{-2mm}{\small Department of Mathematical Sciences} 
\\ \vspace{-2mm}{\small 850 Columbia ave} \\ \vspace{-2mm}{\small Claremont, CA 91711, USA} \\ \vspace{-2mm}{\small phone: 1(909)607-7261} \\ \vspace{-2mm}{\small fax: 1(909)621-8419} 
\\{\small \texttt{dgerman@cmc.edu}} 
}

\date{\today}

\begin{document}

\maketitle

\begin{abstract}
We study a financial model with a non-trivial price impact effect. In
this model we consider the interaction of a large investor trading in
an illiquid security, and a market maker who is quoting prices for
this security. We assume that the market maker quotes the prices such
that by taking the other side of the investor's demand, the market
maker will arrive at maturity with the maximal expected utility of the
terminal wealth. Within this model we provide an explicit recursive pricing formula for an exponential utility function, as well as an asymptotic expansion for the price for a ``small'' simple demand.
\end{abstract}



\newpage
\section{Introduction}
\label{sec:intro}

The study of contingent claim valuation problem accounts for a large number of papers in Finance, Economics and Mathematical Finance in particular. This question was and currently is studied by many authors in various models under different assumptions. A common assumption made by many authors is the basic economic assumption (imposed either implicitly or explicitly) that an economic agent can trade any security in the \emph{desired} quantity at the \emph{same} price. The consequence of this assumption is that the economic agent's actions do not affect the traded security's price, and that there is never a shortage of any security in any quantity. 

One way to relax this assumption is to consider a model where agent's
actions move prices. We will achieve that by introducing the notion of
liquidity into the model. Liquidity is a complex concept standing for
\emph{the ease of trading of a security}.  (Il)liquidity can have
different sources, such as inventory risk -- \cite{Stoll:78},
transaction costs -- \cite{CvitKar:95}, uncertain
holding horizons -- \cite{Huang:03}, asymmetry of information -- \cite{GarPed:04},
demand pressure -- \cite{GarPedPot:09}, search friction -- \cite{DufGarPed:05},
stochastic supply curve -- \cite{CetinJarrowProtter:04} and demand for immediacy -- 
\cite{GrossMiller:88}, among many others (see \cite{AmihudMendPed:05} for a
thorough literature overview).

We will consider the interaction of a large investor trading in an illiquid
security, and a market maker who is quoting prices for this security. We will
assume that the market maker quotes the prices such that by taking the other
side of the investor's demand, she will arrive at maturity with the maximal
expected utility of the terminal wealth.  This idea was also used in a recent
paper \cite{GarPedPot:09} by G\^{a}rleanu, et. al. In Section
\ref{sec:model} we will rigorously define a model for a large
investor. Within this model in this paper we will be concerned with the
following questions: ``Does there exist a price process corresponding to
an arbitrary demand of the large investor, and whether this process is
unique.'' An affirmative answer to these questions (under certain
conditions) is presented in Section \ref{sec:pw_cont_subsec}.

An equally important problem is the replication of
contingent claims in the large investor model with price impact. A
companion paper by \cite{German:10a} shows the existence of a unique pricing rule
for a broad class of derivative securities and utility functions, as well as
the existence of a unique trading strategy that leads to a perfect
replication.

Our approach to the model of a large investor follows the
traditional framework of Economic Theory. We begin with economic primitives
(such as agent's preferences and market equilibrium) and {\it then} derive the
model. This is different from several papers in Mathematical Finance where the
nature of illiquidity is postulated {\it a priori}, see for example
\cite{CvitMa:96}, \cite{CetinJarrowProtter:04}, \cite{BankBaum:04} and
\cite{Frey:98}.

\section{Large investor market model}
\label{sec:model}

We assume that the uncertainty and the flow of information are modeled by a
filtered probability space $(\Omega,\FF, (\FF_t)_{0\le t\le T}, \PP)$, where
the filtration $\FF$ is generated by a $J$-dimensional Brownian Motion $B$,
that is,
\begin{equation}
  \label{eq:5}
  \FF_t = \FF_t^B, \quad 0\leq t\leq T.  
\end{equation}
Here $T$ is a finite time horizon, and $\FF=\FF_T$.

The security market consists of $J$ risky assets and a riskless asset. These
assets are traded between the investor and the market maker. We work in
discounted terms and (without loss of generality) assume that the return on
the riskless asset is zero. We denote by $\FF_T$-measurable random variables
$f=(f^j)_{1\le j\le J}$ the payoffs of the risky assets at maturity and by
$S^H=(S^H_t)_{0\le t\le T}$ the ($J$-dimensional) price process of the risky
assets under the condition that the investor is using the ($J$-dimensional)
trading strategy or \emph{demand process} $H = (H_t)_{0\le t\le T}$. Of
course, at maturity the price does not depend on the strategy:
\begin{equation}\label{eq:price_at_mat} 
  S^H_T = f, \text{ for all } H. 
\end{equation}

From here on we will implicitly understand that we have $J$-dimensional
processes, and without loss of generality we will use one-dimensional
notation.

The market maker can be viewed then as a {\it liquidity provider}. She takes
the other side of the investor's demand, which can be positive, as well as
negative. We assume that the market maker always responds to the investor's
demand, that is the market maker always quotes the price (which turns out to
be a function of the trade size). 
% The reason for this assumption is that the
% market maker is naturally forced to quote the prices to achieve the
% equilibrium by meeting the investor's demand. Of course by equilibrium we mean
% that both parties are ``happy'' with the current prices, have no desire to act
% to change these prices, and the supply is equal to the demand. 
Moreover, the market maker quotes the pricies such that she arrives at
maturity with the maximized expected utility of her terminal wealth.

% In order to
% describe ``happiness'' of the market maker we use the standard apparatus of
% utility functions. We assume that the market maker has a utility function
% $U:\Real\rightarrow\Real$, which is strictly increasing, strictly concave,
% continuously differentiable, and satisfies the Inada conditions
 It may be tempting to think that the market maker would quote positive
  infinity price when the investor is buying, and negative infinity when the
  investor is selling. However, by the natural economic assumption,
  the price process $S$ of the contingent claim $f$ is a semimartingale. Moreover, by 
  \eqref{eq:price_at_mat} the price process at maturity is equal to $f$, and therefore the
  plus/minus infinity price processes are ruled out
  here. 
 
We will use the standard apparatus of
utility functions. We assume that the market maker has a utility function
$U:\Real\rightarrow\Real$, which is strictly increasing, strictly concave,
continuously differentiable, and satisfies the Inada conditions
\begin{align*}
  U'(-\infty)&=\lim_{x\rightarrow-\infty }U'(x)=\infty, \\
  U'(\infty)&=\lim_{x\rightarrow\infty}U'(x)=0.
\end{align*}

We shall also require the following two technical assumptions.
\begin{assumption}\label{as:exp_moments}
  The terminal value of the traded asset $\left.f\!=\!(f^j)_{1\le j\le J}\!\in\FF_T\right.$ has finite
  exponential moments, that is
  \begin{align*}
    \EE[\exp(\langle q,f\rangle)]<\infty, \quad
    q\in\Real^J.
  \end{align*}
\end{assumption}

\begin{assumption}\label{as:exp_utility}
  Utility function $U:\Real\longrightarrow\Real$ satisfies
  \begin{align}\label{eq:exp_utility}
    c_1 < -\frac{U'(x)}{U''(x)} < c_2 \text{ for some } c_1,c_2>0.
  \end{align}
\end{assumption}

\noindent Notice that a linear combination of exponential functions of the form
\begin{align*}
  U(x)=\sum_{i=1}^N-c_i\frac{e^{-\gamma_ix}}{\gamma_i}, \quad \gamma_i,c_i>0,\
  x\in\Real
\end{align*}
satisfies the assumption above. We also notice that Assumption \ref{as:exp_utility} implies the Inada conditions.

We assume that the investor reveals his market orders (his demand process) $H$
to the market maker. The market maker responds to the investor's demand by
quoting the price, and by taking the other side of the demand. That is, if $H$
is the investor's strategy, then $-H$ is the market maker's strategy. In other
words, the market maker responds to the demand so that the market rests in
equilibrium (supply equals demand). The market maker is quoting the price in
such a way that she arrives at maturity with the maximal expected
utility of the terminal wealth. Formally this can be stated as

\begin{definition}
  \label{def:price_proc}
  Let $x\in\Real$ be the initial cash endowment of the \mm. Let $f=(f^j)_{1\le
    j\le J}$ be an $\FF_T$-measurable contingent claim. Let $H=(H^j)_{1\le
    j\le J}$ be a predictable process. The equivalent probability measure
  $\PP^H\sim\nobreakspace\PP$ is called \emph{the pricing measure of $f$ under demand $H$},
  and the semimartingale $S^H$ is called \emph{the price process of $f$
    under demand $H$} if
  \begin{align}
    \label{eq:def:density}
    \frac{d\PP^H}{d\PP}\triangleq \frac{U'(x-\int_0^TH_udS^H_u)}{\EE[U'(x -
      \int_0^TH_udS^H_u)]},
  \end{align}
  and the price process $S^H$ with the integral $\int HdS^H$ are martingales
  under $\mathbb{P}^H$. In particular,
  \begin{align*}
    S^H_t\triangleq\EE^H[f|\FF_t], \quad 0\le t\le T.
  \end{align*}
\end{definition}

The above definition displays an intimate relationship between the price
process and the pricing measure. It may not be clear from the formulation of
Definition \ref{def:price_proc} that it reflects the mechanics of the market
described in the previous paragraph. However, notice that the density of
$\PP^H$ is chosen in such a way that the process $-H$ is indeed a solution to
the \mm's optimization problem (which will be defined below.) Naturally, the
semimartingale $S^H$ is defined in such a way that it is a martingale
under the pricing measure. It will become evident from the following lemma,
that the numerator of \eqref{eq:def:density} is nothing else but the \mm's
marginal utility.

\begin{lemma}\label{lemma:equivalent_def}
  Let $x\in\Real$ be the initial cash endowment of the \mm. Suppose
  $f$ satisfies Assumption \ref{as:exp_moments}, and $U$ satisfies
  Assumption \ref{as:exp_utility}. Let $H=(H^j)_{1\le j\le J}$ be a
  predictable process. Suppose that $S^H$ is the price process of $f$
  under demand $H$. Then $-H$ is the unique solution of the
  optimization problem
  \begin{align}\label{eq:u(x)}
    u(x)\triangleq\max_{G\in\HH(S^H, \PP^H)}\EE[U(x+\int_0^TG_udS^H_u)],
  \end{align}
  where $\HH(S^H,\PP^H)$ is the collection of predictable processes $G$ such
  that $$\int_0^{\cdot}G_udS^H_u$$ is a $\PP^H$-martingale.
\end{lemma}

The proof of this Lemma is given in the companion paper \cite{German:10a}.

In the following section we will be interested in finding the answers to the following
questions: 
\begin{itemize}
  \item Does the price process $S^H$ exist for an arbitrary demand $H$?
  \item Provided that $S^H$ exists, is it unique?
\end{itemize}

%%%%%%%%%%%%%%%%%%%%%%%%%%%%%%%%%%%%%%%%%%%%%%%%%%
\section{Price process under simple demand and exponential
  utility}\label{sec:pw_cont_subsec}

In our model the market maker has to meet the demand $H$, which forces her
strategy to be $-H$. Therefore the value of the market maker's portfolio at
time $t$ is $x-\int_0^tH_udS^H_u$, when $S^H$ is the price that the market
maker is quoting depending on the demand $H$, and $x$ is the market maker's
initial wealth. An important question is whether for every predictable demand
process $H$ and an $\FF_T$-measurable random variable $f$ there is a
corresponding price process $S^H$, that satisfies Definition
\ref{def:price_proc}.

In the following theorem we show that the price process $S^H$ exists and is
unique for an exponential utility $U$ and bounded \emph{simple} demand processes
$H$.  Moreover, we provide an explicit recursive algorithm allowing the
computation of $S^H$.  We start by recalling the definition of simple
strategies.

\begin{definition}\label{def:simple_proc}
  Let $(t_i)_{0\le i \le N}$ be a partition of the time interval $[0,T]$ into
  $N$ intervals with $t_0=0$, and $t_{N}=T>0$. Let $\{\theta^i\}_{0\le i\le
    N-1}$ be a sequence of $(\FF_{t_i})_{0\le i\le N-1}$-measurable
  $J$-dimensional random variables respectively. Then the process
  \begin{equation}
    \label{eq:simple_demand}
    H_t =\theta^0\ind_{\{0\}}(t)+ \sum_{i=0}^{N-1}\theta^i\ind_{(t_i, t_{i+1}]}(t)
  \end{equation}
  is called \emph {a simple process}.
\end{definition}
 
We can now formulate the main results of this section.

\begin{theorem}\label{thm:recursive_price}
  Assume that the utility function $U$ is of exponential form:
  \begin{displaymath}
    U(x)=-\frac{1}{\gamma}e^{-\gamma x}, \gamma>0,
  \end{displaymath}
  $f$ is bounded, and \eqref{eq:5} holds true. Then for a bounded simple
  demand process $H$ given by \eqref{eq:simple_demand} the price process $S^H$
  exists and is unique. Moreover, it satisfies the following equations of
  backward induction:
  \begin{equation}
    \label{eq:rec_price}
    \begin{split}
      & S_T^H = f, \\
      &S^H_t = \frac{\EE\left[\left.f\exp\left\{\gamma
              \left(\theta^kS^H_{t_{k+1}}+\sum_{i=k+1}^{N-1}
                \theta^i(S^H_{t_{i+1}}-S^H_{t_i})\right)
            \right\}\right|\FF_t\right]}{\EE\left[\left. \exp\left\{
              \gamma\left(\theta^kS^H_{t_{k+1}}+
                \sum_{i=k+1}^{N-1}\theta^i(S^H_{t_{i+1}}-S^H_{t_i})\right)
            \right\}\right|\FF_t\right]},
      \\
      &\hbox{ for } t\in[t_k,t_{k+1}].
    \end{split}
  \end{equation}
  The pricing measure $\mathbb{P}^H$ is given by
  \begin{equation}
    \label{eq:15}
    \frac{d\PP^H}{d\PP} =
    \frac{\exp\{\gamma\sum_{i=0}^{N-1}\theta^i(S^H_{t_{i+1}}-S^H_{t_i})\}}{\EE\left[
        \exp\{\gamma\sum_{i=0}^{N-1}\theta^i(S^H_{t_{i+1}}-S^H_{t_i})\}\right]}. 
  \end{equation}
\end{theorem}

\begin{remark}
  Theorem \ref{thm:recursive_price} was formulated in \cite{GarPedPot:09} (Theorem 1) in a different form and without a formal proof. For the sake of mathematical completeness we here provide a proof.
\end{remark}

\begin{proof}
  We first observe that since both $f$ and $H$ are bounded, the process $S^H$
  is also bounded and well defined. Therefore the density of $\PP^H$, which is
  defined in terms of the process $S^H$, is well defined.

  In this proof we will show that $S^H$ and $\PP^H$ defined by
  \eqref{eq:rec_price} and \eqref{eq:15} satisfy Definition
  \ref{def:price_proc}.

  For the exponential utility function, \eqref{eq:15} can be written as
  \begin{align*}
    \frac{d\PP^H}{d\PP} =
    \frac{\exp\{\gamma\sum_{i=k+1}^{N-1}\theta^i(S^H_{t_{i+1}}-S^H_{t_i})\}}{
      \EE\left[\exp\{\gamma\sum_{i=k+1}^{N-1}\theta^i(S^H_{t_{i+1}}-S^H_{t_i})\}\right]}
    = \frac{U'(x-\int_0^TH_udS^H_u)}{\EE[U'(x-\int_0^TH_udS^H_u)]},
  \end{align*}
  for any $x\in\Real$.

  Next we will verify by backward induction that $S^H$ defined by
  \eqref{eq:rec_price} can be also represented as
  $S^H_t=\EE^H[f|\FF_t]$. Indeed, let $t\in[t_{N-1},T]$.  Since
  \begin{align*}
    \sum_{i=0}^{N-2}\theta^i(S^H_{t_{i+1}}-S^H_{t_i})-\theta^{N-1}S^H_{t_{N-1}}
  \end{align*}
  is an $\FF_{t_{N-1}}$-measurable random variable,
  \begin{align}
    \EE^H[f|\FF_t]&=\frac{\Ec{f\exp\left\{\gamma\left(
            \sum_{i=0}^{N-1}\theta^i(S^H_{t_{i+1}}-S^H_{t_i})\right)\right\}}{t}}{
      \Ec{\exp\left\{
          \gamma\left(\sum_{i=0}^{N-1}\theta^i(S^H_{t_{i+1}}-S^H_{t_i})
          \right)\right\}}{t}}\nonumber\\
    & =\frac{\exp\left\{\gamma\left(
          \sum_{i=0}^{N-2}\theta^i(S^H_{t_{i+1}}-S^H_{t_i})-\theta^{N-1}S^H_{t_{N-1}}
        \right)\right\}}{ \exp\left\{\gamma\left(
          \sum_{i=0}^{N-2}\theta^i(S^H_{t_{i+1}}-S^H_{t_i})-\theta^{N-1}S^H_{t_{N-1}}
        \right)\right\}}\nonumber\\
    &\quad\times\frac{\Ec{fe^{\gamma\theta^{N-1}f}}{t}}{\Ec{
        e^{\gamma\theta^{N-1}f}}{t}}\nonumber\\
    &=\frac{\Ec{fe^{\gamma\theta^{N-1}f}}{t}}{\Ec{e^{\gamma\theta^{N-1}f}}{t}}.
    \label{eq:rec_n=N-1}
  \end{align}
  The equality \eqref{eq:rec_n=N-1} shows us that on the time interval
  $t\in[t_{k-1},t_k]$ with $k=N-1$, $S^H_t$ given by \eqref{eq:rec_price} is
  in fact equal to $\EE^H[f|\FF_t]$.

  Assume now that $S^H_t=\EE^H[f|\FF_t]$ for $t\in[t_{k+1},t_{k+2}]$. Let
  $t\in[t_{k},t_{k+1}]$. Since
  \begin{align*}
    \sum_{i=0}^{k-1}\theta^i(S^H_{t_{i+1}}-S^H_{t_i})-\theta^kS^H_{t_k}
  \end{align*}
  is an $\FF_{t_k}$-measurable random variable, by induction
  \begin{align}
    \EE^H[f|\FF_t]&=\frac{\Ec{f\exp\left\{\gamma\left(
            \sum_{i=0}^{N-1}\theta^i(S^H_{t_{i+1}}-S^H_{t_i})
          \right)\right\}}{t}}{\Ec{\exp\left\{\gamma\left(
            \sum_{i=0}^{N-1}\theta^i(S^H_{t_{i+1}}-S^H_{t_i})\right)\right\}}{t}}
    \nonumber\\
    &
    =\frac{\exp\left\{\gamma\left(\sum_{i=0}^{k-1}\theta^i(S^H_{t_{i+1}}-S^H_{t_i})-
          \theta^kS^H_{t_k}\right)\right\}}{\exp\left\{
        \gamma\left(\sum_{i=0}^{k-1}\theta^i(S^H_{t_{i+1}}-S^H_{t_i})-
          \theta^kS^H_{t_k}\right)\right\}}\nonumber\\
    &\quad\times\frac{\Ec{f\exp\left\{\gamma\left(\theta^kS^H_{t_{k+1}}+
            \sum_{i=n+1}^{N-1}\theta^i(S^H_{t_{i+1}}-S^H_{t_i})\right)
        \right\}}{t}}{\Ec{\exp\left\{\gamma\left(\theta^kS^H_{t_{k+1}}+
            \sum_{i=n+1}^{N-1}\theta^i(S^H_{t_{i+1}}-S^H_{t_i})\right)\right\}}{t}}
    \nonumber\\
    &=\frac{\Ec{f\exp\left\{\gamma\left(\theta^kS^H_{t_{k+1}}+\sum_{i=n+1}^{N-1}
            \theta^i(S^H_{t_{i+1}}-S^H_{t_i})\right)\right\}}{t}}{\Ec{
        \exp\left\{\gamma\left(\theta_kS^H_{t_{k+1}}+\sum_{i=n+1}^{N-1}
            \theta^i(S^H_{t_{i+1}}-S^H_{t_i})\right)\right\}}{t}}.\label{eq:check_Sh}
  \end{align}
  Now we can see that \eqref{eq:check_Sh} is precisely equal to
  \eqref{eq:rec_price}. Hence we proved that $S^H$ defined by
  \eqref{eq:rec_price} is indeed the price process of $f$ under demand
  $H$. Uniqueness follows from the construction.
\end{proof}

\begin{example}\label{example:bachelier}
  Consider a Bachelier model with $f=\zeta B_T$, constant demand
  $H_t=\theta$, and $U(x)=-\frac{e^{-\gamma x}}{\gamma}$, where
  $\theta,\gamma,\zeta\in\Real, \ 0<\gamma<\infty$. $B$ is a
  one-dimensional Brownian Motion under $\PP$.

  Although in this example we would like to illustrate an application
  of Theorem \ref{thm:recursive_price}, for technical reasons we chose $f$ to be an
  unbounded random variable (which does not satisfy the conditions of
  the theorem.) Therefore we cannot directly apply the result of the
  above theorem. However, instead we will refer to Definition
  \ref{def:price_proc}, and in this simple case we can work out the
  price process right from the definition.

  By the Definition \ref{def:price_proc}, the pricing measure $\PP^H$
  is given by
  \begin{equation}\label{eq:example:sh}
    \begin{split}
      \frac{d\PP^H}{d\PP}&=\frac{U'(x-\int_0^TH_udS^H_u)}{\EE[U'(x-\int_0^TH_udS^H_u)]}
      =
      \frac{\exp\{\gamma\int_0^TH_udS^H_u\}}{\EE[\exp\{\gamma\int_0^TH_udS^H_u\}]}\\
      &= \frac{\exp\{\gamma\theta(S^H_T-S^H_0)\}}{
        \EE[\exp\{\gamma\theta(S^H_T-S^H_0)\}]},
    \end{split}
  \end{equation}
  where the last equality follows from the fact that $H=\theta$ is a
  constant.  We notice that since $S_T^H=f=\zeta B_T$, and $S_0^H$ is
  a constant, \eqref{eq:example:sh} can be written as
  \begin{align*}
    \frac{d\PP^H}{d\PP}=\frac{\exp\{\gamma\theta(f-S^H_0)\}}{
      \EE[\exp\{\gamma\theta(f-S^H_0)\}]}=\exp\left\{\gamma\theta\zeta
      B_T-\half (\gamma\theta\zeta)^2T\right\}.
  \end{align*}
  Girsanov's Theorem implies that under the probability measure
  $\PP^H$ there exists a Brownian Motion $B^H$ such that
  \begin{align*}
    B^H_t=B_t-\gamma\theta\zeta t,
  \end{align*}
  and therefore since $S^H$ is the price process of $\zeta B_T$ under
  demand $\theta$, Definition \ref{def:price_proc} implies that
  \begin{equation}
    \begin{split}
      S^H_t&=\EE^H[f|\FF_t]=\EE^H[\zeta B_T|\FF_t]=\EE^H[ \zeta
      (B^H_T+\gamma\theta\zeta T)|\FF_t]=\zeta
      B^H_t+\gamma\theta\zeta^2
      Tr\\
      &=\zeta B_t+\gamma\theta\zeta^2(T-t)= S^0_t +
      \gamma\int_t^TH_u\zeta^2du, \
      t\in[0,T], \label{eq:const_dem_price}
    \end{split}
  \end{equation}
where
\begin{equation*}
  S^0_t=\zeta B_t=\Ec{f}{t}.
\end{equation*}
\end{example}

Theorem \ref{thm:recursive_price} provides a unique price process $S^H$ of the
contingent claim $f$ under demand $H$.  Unfortunately, $S^H$ is computed in a
recursive form, which makes its practical use rather limited. The
following theorem gives a convenient asymptotic expansion for $S^H$ in the
case of a ``small'' simple demand $\epsilon H$.

\begin{theorem}
  \label{thm:expansion_theorem}
  % In addition to conditions of Theorem \ref{thm:recursive_price}, assume
  % that $f$ is a bounded random variable. Assume that the demand process $H$
  % is simple and bounded, and the utility function is exponential
  % \begin{displaymath}
  %   U(x)=-\frac{e^{-\gamma x}}{\gamma}, \quad x\in \mathcal{R} 
  % \end{displaymath}
  % for some $\gamma>0$.
  Assume that conditions of Theorem \ref{thm:recursive_price} hold true.  Then
  for $\epsilon>0$ we have
  \begin{equation}
    \label{eq:SeHstatement}
    \SeH_t = S^0_t + \epsilon\EE\left[\left.\int_t^TH_ud\langle
        S^0\rangle_u\right|\FF_t\right] + \xi_t(\epsilon), \quad t\in[0,T], 
  \end{equation}
  where
  \begin{equation}
    \label{eq:17}
    S^0_t=\EE[f|\FF_t], \quad t\in[0,T],
  \end{equation}
  and
  \begin{equation}
    \label{eq:limitstatement}
    \lim_{\epsilon\rightarrow 0} \frac{\xi_t(\epsilon)}{\epsilon}=0
    \text{ for any } t\in[0,T], 
  \end{equation}
  where the convergence is in probability.
\end{theorem}

\begin{remark}
  Theorem \ref{thm:expansion_theorem} is qualitatively rather similar to Theorem 3 in \cite{GarPedPot:09}, but it offers the convenience of not being recursive, unlike Theorem 3 in \cite{GarPedPot:09}.
\end{remark}

\begin{proof}
  We will proceed by backward induction. According to Theorem
  \ref{thm:recursive_price}, the price process $\SeH$ under demand $\epsilon
  H$, it is equal to
  \begin{align*}
    &\SeH_t=\frac{\EE\left[\left.f\exp\left\{\epsilon\gamma
            \left(\theta^k\SeH_{t_{k+1}}+\displaystyle\sum_{i=k+1}^{N-1}
              \theta^i(\SeH_{t_{i+1}}-
              \SeH_{t_i})\right)\right\}\right|\FF_t\right]}{
      \EE\left[\left.\exp\left\{ \epsilon\gamma\left(\theta^k\SeH_{t_{k+1}}+
              \displaystyle \sum_{i=k+1}^{N-1}
              \theta^i(\SeH_{t_{i+1}}-\SeH_{t_i})\right)
          \right\}\right|\FF_t\right]},\\
    &\hbox{ for } t\in[t_k,t_{k+1}].
  \end{align*}
  In particular, for $t\in[t_{N-1},T]$
  \begin{align}\label{eq:first_iteration}
    \SeH_t=\frac{\Ec{fe^{\epsilon\gamma\theta^{N-1}f}}{t}}{
      \Ec{e^{\epsilon\gamma\theta^{N-1}f}}{t}},
  \end{align}
  and therefore
  \begin{align}
    \evalepszero{\SeH_t}=\Ec{f}{t}=S^0_t.\label{eq:Szero}
  \end{align}
  Expression (\ref{eq:Szero}) gives economical meaning to the process
  $S^0$. It is a price process under zero demand, which is a $\PP$-martingale.

  Notice that since $f$ and $\theta^{N-1}$ are bounded, the Dominated
  Convergence Theorem implies that
  \begin{align*}
    &\partialepsilon\Ec{e^{\epsilon\gamma\theta^{N-1}f}}{t}\\
    &\quad=\lim_{\delta\rightarrow
      0}\frac{\Ec{e^{(\epsilon+\delta)\gamma\theta^{N-1}f} -
        e^{\epsilon\gamma\theta^{N-1}f}}{t}}{\delta}
    =\Ec{\gamma\theta^{N-1}e^{\epsilon\gamma\theta f}}{t},
  \end{align*}
  as well as
  \begin{align*}
    \partialepsilon\Ec{fe^{\epsilon\gamma\theta^{N-1}f}}{t}
    =\Ec{\gamma\theta^{N-1}fe^{\epsilon\gamma\theta^{N-1} f}}{t}.
  \end{align*}
  Hence
  \begin{align}
    \partialepsilon\SeH_t&=\partialepsilon
    \frac{\Ec{fe^{\epsilon\gamma\theta^{N-1}f}}{t}}{
      \Ec{e^{\epsilon\gamma\theta^{N-1}f}}{t}}\nonumber\\
    &=\frac{\Ec{\gamma\theta^{N-1}f^2 e^{\epsilon\gamma\theta^{N-1}f}}{t}\Ec{
        e^{\epsilon\gamma\theta^{N-1}f}}{t}}{\Ec{
        e^{\epsilon\gamma\theta^{N-1}f}}{t}^2}
    \nonumber\\
    &\quad- \frac{\Ec{\gamma\theta^{N-1}fe^{\epsilon\gamma\theta^{N-1}f}}{t}
      \Ec{fe^{\epsilon\gamma\theta^{N-1}f}}{t}}{
      \Ec{e^{\epsilon\gamma\theta^{N-1}f}}{t}^2},\quad
    t\in[t_{N-1},T]. \label{eq:diffSeH}
  \end{align}
  We notice that for every $\epsilon$, \eqref{eq:diffSeH} is a finite random
  variable.  It follows from (\ref{eq:diffSeH}) that
  \begin{align*}
    \evalepszero{\partialepsilon\SeH_t}=\gamma\theta^{N-1}( \Ec{f^2}{t}-
    \Ec{f}{t}^2)=\Ec{\gamma\int_t^TH_ud\variance{S^0}_u}{t},
  \end{align*}
  for $t\in[t_{N-1},T]$. Therefore we can compute Taylor's expansion of $\SeH$
  around zero on $[t_{N-1},T]$:
  \begin{align*}
    \SeH_t=\frac{\Ec{fe^{\epsilon\gamma\theta^{N-1}f}}{t}}{
      \Ec{e^{\epsilon\gamma\theta^{N-1}f}}{t}}=\Ec{f}{t}
    +\epsilon\Ec{\gamma\int_t^TH_ud\variance{S^0}_u}{t}+\xi_t(\epsilon),
  \end{align*}
  where $\frac{\xi_t(\epsilon)}{\epsilon}$ converges to $0$ in probability as
  $\epsilon\to 0$. This proves the assertion of the Theorem on the interval
  $[t_{N-1},T]$.

  Now let us assume that for $t\in[t_{m+1},T]$
  \begin{equation}
    \SeH_t = S^0_t + \epsilon\EE\left[\left.\gamma\int_t^TH_ud\langle
        S^0\rangle_u\right|\FF_t\right] + \xi_t(\epsilon).
    \label{eq:induction_SeH} 
  \end{equation}
  Then for $t\in[t_{m},T]$
  \begin{align*}
    \partialepsilon\SeH_t &=\partialepsilon
    \frac{\Ec{f\exp\left\{\epsilon\gamma\brakets{\theta^m
            \SeH_{t_{m+1}}+\displaystyle\sum_{i=m+1}^{N-1}\theta^i(\SeH_{t_{i+1}}
            -\SeH_{t_i})}\right\}}{t}}{\Ec{\exp\left\{
          \epsilon\gamma\brakets{\theta^m\SeH_{t_{m+1}}
            +\displaystyle\sum_{i=m+1}^{N-1}
            \theta^i(\SeH_{t_{i+1}}-\SeH_{t_i})}\right\}}{t}}\\
    &=\partialepsilon\frac{\Ec{fL^m}{t}}{\Ec{L^m}{t}}
  \end{align*}
  where
  \begin{displaymath}
    L^m=\exp\left\{\epsilon\gamma\brakets{\theta^m\SeH_{t_{m+1}}+
        \displaystyle\sum_{i=m+1}^{N-1}\theta^i(\SeH_{t_{i+1}}
        -\SeH_{t_i})}\right\}.
  \end{displaymath}
  
  We have
  \begin{displaymath}
    L^m|_{\epsilon = 0} = 1
  \end{displaymath}
  and
  \begin{displaymath}
    \frac{\partial}{\partial \epsilon} L^m|_{\epsilon=0} =
    \gamma\brakets{\theta^m S^0_{t_{m+1}}+
      \displaystyle\sum_{i=m+1}^{N-1}\theta^i(S^0_{t_{i+1}}
      - S^0_{t_i})}. 
  \end{displaymath}

  It follows that
  \begin{align*}
    &\evalepszero{\partialepsilon\SeH_t}
    =\gamma\Ec{f\left(\theta^mS^0_{t_{m+1}}+
        \displaystyle\sum_{i=m+1}^{N-1}\theta^i
        (S^0_{t_{i+1}}-S^0_{t_i})\right)}{t}\\
    &\quad-\gamma\Ec{\theta^mS^0_{t_{m+1}}+\displaystyle
      \sum_{i=m+1}^{N-1}\theta^i(S^0_{t_{i+1}}-S^0_{t_i})}{t}\Ec{f}{t}\\
    &=\gamma\Ec{f\brakets{\int_{t_m}^TH_udS^0_u+\theta^mS^0_{t_m}}}{t}\\
    &\quad-\gamma\Ec{\int_{t_m}^TH_udS^0_u+\theta^mS^0_{t_m}}{t}\Ec{f}{t}\\
    &=\gamma\left(\int_0^tdS^0_u\int_0^tH_udS^0_u +
      \Ec{\int_t^TH_ud\variance{S^0}_u}{t} + S^0_0\int_0^tH_udS^0_u \right.\\
    &\quad - S^0_t\int_0^{t_m}H_udS^0_u
    - \left.S^0_t\int_0^{t}H_udS^0_u + S^0_t\int_0^{t_m}H_udS^0_u \right)\\
    &=\gamma\Ec{\int_t^TH_ud\variance{S^0}_u}{t}, \quad t\in[t_m,T].
  \end{align*}
  Hence, Taylor's expansion on the time interval $[t_m, T]$ is given by
  \begin{align*}
    \SeH_t=\Ec{f}{t}+\epsilon\Ec{\gamma\int_t^TH_ud\variance{S^0}_u}{t}
    +\xi_t(\epsilon),\ t\in[t_m, T]
  \end{align*}
  where $\frac{\xi_t(\epsilon)}{\epsilon}$ converges to $0$ in probability as
  $\epsilon\to 0$.
\end{proof}

\begin{remark}
  Note that in the framework of Example \ref{example:bachelier} the first
  order expansion \eqref{eq:SeHstatement} is exact, that is, in this case the
  ``error'' term $\xi_t(\epsilon)=0$.
\end{remark}

\section{Conclusion}

We studied the problem of pricing of an illiquid asset in the model
with price impact. We derived a recursive unique pricing rule for the
illiquid asset under the conditions that the market maker's utility
function is exponential, the asset is bounded, and the demand is
piece-wise constant. The result was proved by construction.

We also derived the asymptotic expansion \eqref{eq:SeHstatement}, which gives a very
convenient expression for the price process under a ``small'' simple
demand. Its generalization to continuous demand processes presents an interesting future research project.

\newpage
\bibliographystyle{alpha}
\bibliography{../../../AUX/finance}
\end{document}